# A global assessment of tourism and recreation conservation threats to prioritise interventions

David Lusseau[1] & Francesca Mancini
University of Aberdeen, School of Biological Sciences, Aberdeen AB24 2TZ, UK
[1]d.lusseau@abdn.ac.uk

**Abstract**
We are increasingly using nature for tourism and recreation, an economic sector now generating more than 10% of the global GDP and 10% of global total employment. This growth though has come at a cost and we now have 5930 species for which tourism and recreation are conservation threats. For the first time we use global social media data to estimate where people go to experience nature and determine how this tourism and recreation pressure overlap with the distribution of threatened species. The more people seek interactions with nature in an area, the larger the number of species threatened by those interactions is. Clear crisis areas emerge where many species sensitive to tourism are exposed to high tourism pressures and those are mainly coastal marine regions. Our current tourism management approaches are not achieving biodiversity conservation. The global increase in nature tourism and recreation is set to continue and we need a global consistent response to mitigate its biodiversity impact. Like with other extractive industries, we must prioritise our efforts to diverge tourism away from crisis areas.

**Introduction**
People interact with nature in rich and diverse ways. We extract goods from our natural world, such as timber and meat, but we also use it indirectly by developing our infrastructure or for cultural benefits (Clayton et al., 2017). For example, we seek interactions with nature for leisure; either near our home (recreation) or by travelling away from home (tourism) (Buckley, 2012). Tourism has become one of the world's largest economic sector, accounting for 11% of global GDP and 10% of total employment (World Tourism Organisation, 2017). Nature-based and wildlife tourism have particularly grown faster than the rest of the industry over the past 10 years (Balmford et al., 2009; World Tourism Organisation, 2014). The sustainability of nature-based tourism, and recreation, is often assumed without being challenged despite for example its contribution to the global carbon emission budget (Lenzen et al., 2018) or local extirpation of natural habitats (Gössling, 2002). Indeed, the sector is seen as a key tool in sustainable development that can provide both biodiversity and livelihoods gains (World Bank, n.d.). Nature tourism is not only growing but also changing. While nature-tourism was before perhaps more specialised and more concentrated in protected areas (PA), we now see a decline in PA use combined with a growth of the sector (Balmford et al., 2009; Pergams & Zaradic, 2008). This diversification of nature tourism increases the difficulty to monitor and manage its potential impact on biodiversity and indeed other sustainability targets.

It is now well established that tourism and recreation activities can threaten the survival of species of all taxa (Figure 1). This impact can be simply caused by habitat modification (IUCN Red List threat 1.3, (Pickering & Hill, 2007; Rankin, Ballantyne, & Pickering, 2015)) but also by repeated exposure to disturbance and temporary habitat alteration (IUCN Red List threat 6.1, (Amo, López, & Martín, 2006; Ellenberg, Setiawan, Cree, Houston, & Seddon, 2007; Lusseau, Slooten, & Currey, 2006)). Tourism and recreation is a growing conservation threat and we now have close to 6000 species on the IUCN Red List threatened by these activities. Our understanding of tourism threats is fragmented, coming from many case studies using different methodologies (Senigaglia et al., 2016), and we still do not have a comprehensive global view of the biodiversity footprint of tourism and recreation. This is crucial to understand whether the sector has a global biodiversity impact and whether we need to prioritise regional tourism management to ensure sustainability (Pirotta & Lusseau, 2015).

One key hurdle for such global assessment is that we do not have a global understanding of where people go to experience nature. Recent advances in social media analysis now provide a tool for global estimation of tourism and recreation pressures beyond traditional surveys in tourism hotspots (Hausmann, Toivonen, Heikinheimo, et al., 2017; Hausmann, Toivonen, Slotow, et al., 2017; Francesca Mancini, Coghill, & Lusseau, 2016; Wood, Guerry, Silver, & Lacayo, 2013). Here for the first time we use the largest curated and quality-controlled global open social media dataset to estimate where people interact with nature (Thomee et al., 2016). We estimate the global richness distribution of species threatened by tourism and recreation and determine how the current tourism and recreation pressure overlap with those species. Finally, we assess whether increased pressure is associated with an increased number of species threatened. We determine whether some regions are facing a tourism biodiversity crisis and project whether others are likely to face it in the near future to assess the scale at which the tourism and recreation biodiversity threat should be challenged.

**Methods**

We used the IUCN Red List information to select species threatened by tourism and recreation (Red List threats 1.3 and 6.1) and download the identity of those species from the IUCN Red List website (http://www.iucnredlist.org/search/). The spatial range of species for which it was available (4759 out of 5930) was kindly provided as a direct download by the IUCN Global Species Programme. For others we used GBIF via rgbif (Chamberlain, Boettiger, Ram, Barve, & Mcglinn, 2016) to infer range from sightings other than fossil records and material samples. Only 241 species (4%) remained spatially unassigned due to a lack of spatial information. For all 5930 species we used rredlist to retrieve their habitat type and conservation status and the years in which this status was assessed (Chamberlain, 2017).

We used the YFCC100m Flickr dataset to determine where people use nature (Thomee et al., 2016). This social medium has been validated in a number of studies showing that it is a good proxy human occupancy variable for nature visitation at scales down to 10km in regions with many users (Francesca Mancini et al., 2016). YFCC100m is the most comprehensive publicly available dataset for Flickr, representing roughly 10% of photos available on the platform sampled from 2002 to 2014 with spatial and temporal representation in mind (stratified sampling). We categorised the 100 million photos using the autotag generated by a convoluted neural network model based on photo composition (Thomee et al., 2016) as either a photo of 'nature' or not. Out of the 48,469,829 autotagged and georeferenced photos available, 12,935,520 were tagged as photos of nature (Figure S1).

We gridded the world using hexagonal equal area cells to understand the spatial distribution of both nature-based recreation and tourism as well as species threatened by those activities. We used dggridR to produce a grid of 196,832 $2591km^2$ hexagonal cells as a trade-off among sensible scales for our variables (Barnes, 2017). We used dggridR to count in each cell the number of species present threatened by tourism and recreation as well as the number of nature photos and the total number of photos taken. We therefore could derive for each cell the proportion of nature photos taken out of all photos taken ($p_{nature}$).

We assessed whether the number of species threatened in an area was associated with tourism and recreation intensity using general linear models using generalised least squares (lm-gls). Models were fitted using nlme in R (Pinheiro, Bates, DebRoy, Sarkar, & R Core Team, 2018). We determined whether the $log_{10}$-transformed species count in each cell was associated to the proportion of nature photos ($p_{nature}$) in that cell, accounting for spatial autocorrelation in the residuals using an exponential correlation structure (selected as best representing the autocorrelation structure by model selection). While $p_{nature}$ captures whether a cell receives more attention to its nature by visitors than others, it does not capture the intensity of visitation. To address this question, while avoiding potential collinearity issues, we estimated the lm-gls model for four subset of cells selected depending on the number of nature photos taken in that cell. We categorised cells in the four quartiles of the distribution of $log_{10}$-number of nature photos to capture increased visitation intensity and fitted the same lm-gls to each subset.

Finally, we estimated a measure of overlap between tourism and recreation pressure and threatened biodiversity using a simple relative concordance estimate as the product of the three variables (number of threatened species, number of nature photos, and proportion of nature photos) scaled to their min-max to range from 0 to 1.

**Results**

The number of Red List species threatened by tourism and recreation has increased rapidly over the past 20 years (Figure 1; IUCN threat 1.3: 4310 species, IUCN threat 6.1: 3264 species). In addition, 1858 of those species were assessed multiple times since 1996 and in 94.3% of instances tourism and recreation were only present as a threat in their latter assessment. Most species are not in critical danger (listed data deficient, least concern, or vulnerable; Figure 1) and therefore there is still scope to plan for novel approaches to the global management of tourism and recreation impacts. The most affected habitats are forests in terrestrial biomes and reefs in marine biomes (Figure 1). This is also reflected in the taxonomic distribution of threatened species with anthozoa and plants being more represented (Figure 1).

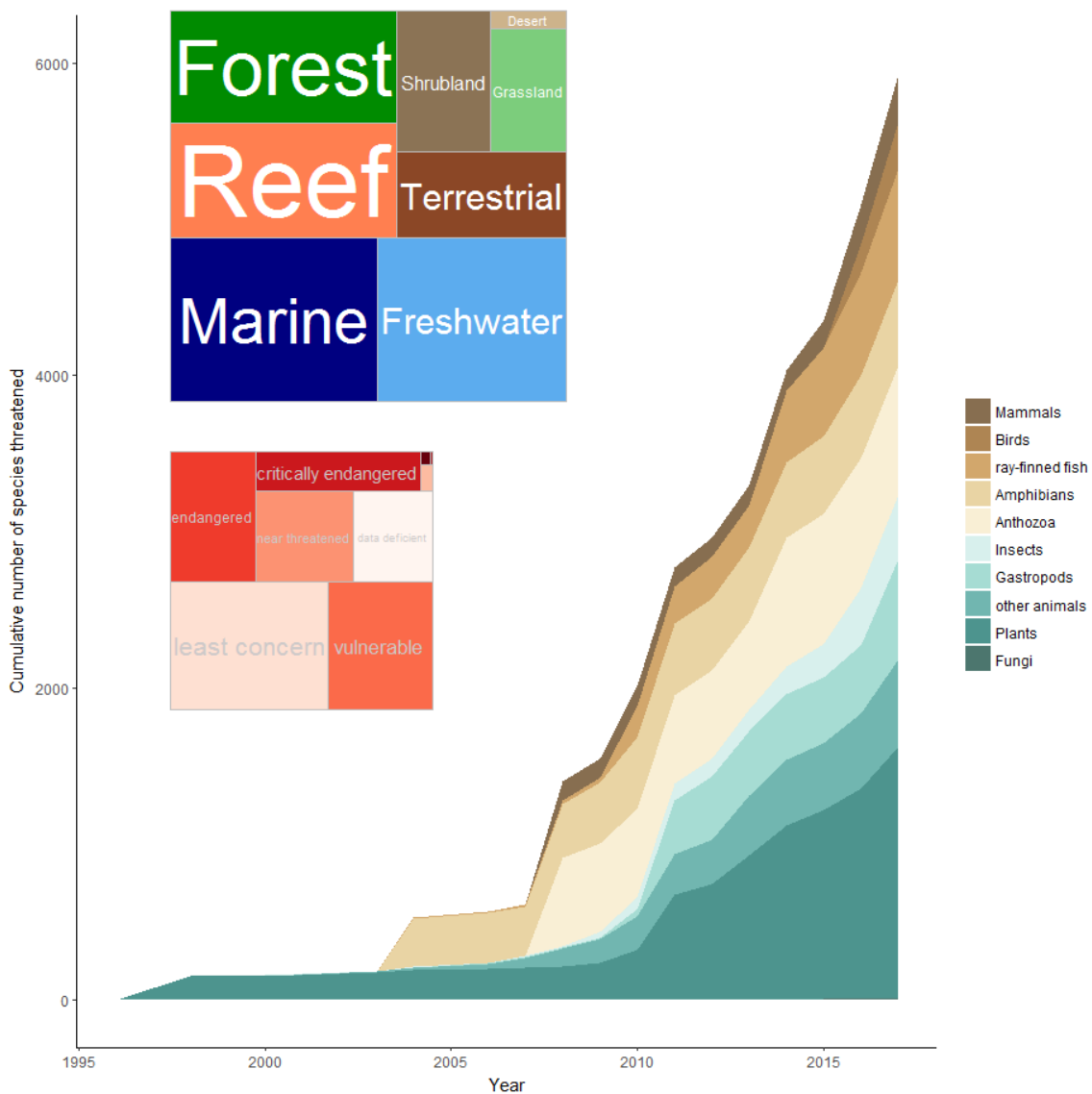

**Figure 1**. Change in the number of species assessed to be threatened by tourism and recreation from 1996 to 2017 including the categorisation of those species by taxon, habitat (inset, relative representation (Tennekes & Ellis, 2017)), and Red List Status (inset, relative representation).

The distribution of threatened species richness (species count per cell) follows a bimodal distribution, with regions with some species threatened by tourism and recreation (about 12 species typically, Figure 2) and others with many threatened species (about 417 species typically, Figure 2). We interpret this latter mode as hotspots of tourism and recreation threat potential. These tend to be coastal marine regions (Figure 2). Some overlap with well-known and mature global marine destinations (eg, Maldives, Australian Great Barrier Reef, Red Sea, Figure 2) while others do not (eg, Somalia).

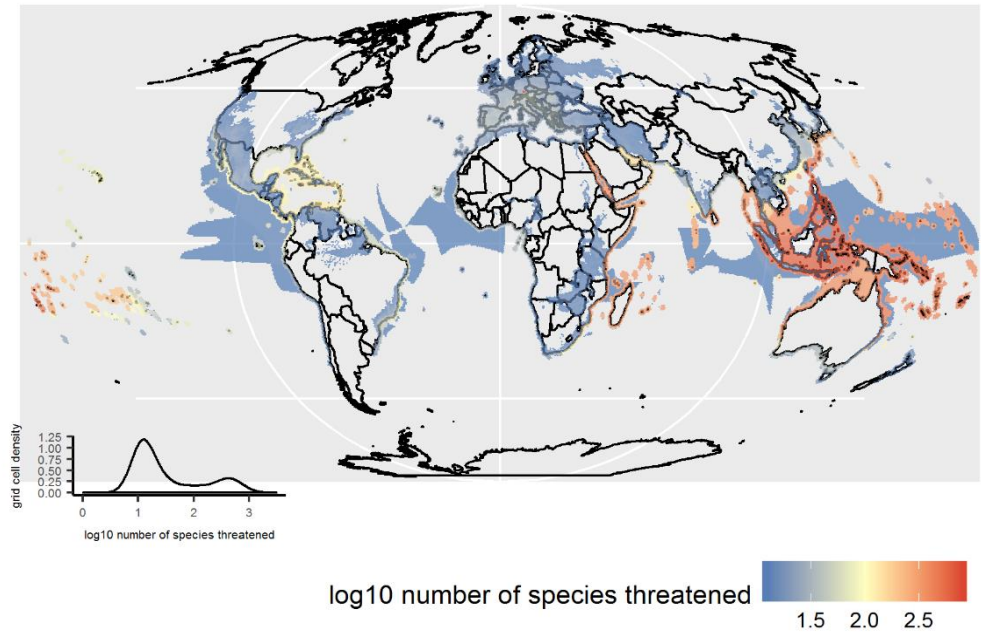

**Figure 2.** The distribution of species threatened by tourism and recreation: number of species ($\log_{10}$) present in each 50 x 50 km grid cells. The distribution is bimodal (inset).

We obtained a spatially explicit global estimate of nature visitation (Figure 3) thanks to the YFCC100m data which can help us distinguish present threat hotspots from threat potential highlighted by the spatial aggregation of threatened species (Figure 2). We find here the well-established nature tourism regions (eg, Western USA, Andes, South and East sub-Saharan Africa, Australia, northern Europe; Figure 3) which differ from the regions more photographed overall (Figure S1). Therefore, accounting for $p_{nature}$ within the context of the total number of nature photos, as previously stated (van Zanten et al., 2016), is a mean to determine the focus on nature an area receives in the context of the intensity of its visitation.

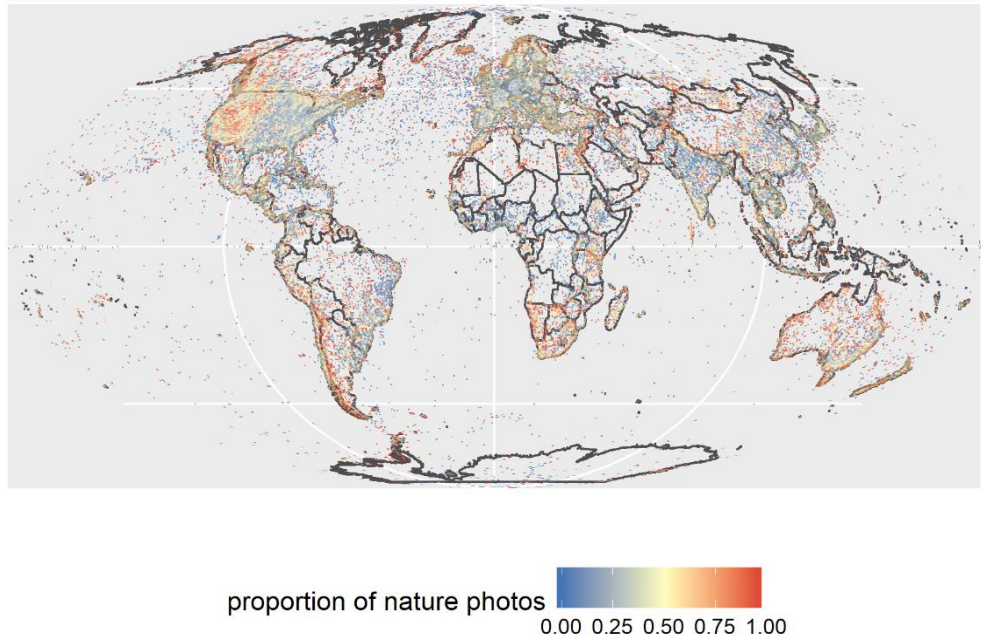

proportion of nature photos
0.00 0.25 0.50 0.75 1.00

**Figure 3.** Estimated distribution of global nature visitation: the proportion of photos posted on Flickr in each 50 x 50km grid cells categorised as nature photos in relation to the total number of photos taken in each grid cells.

We find indeed that the more focus on nature there is in a grid cell, the greater the number of species threatened by tourism and recreation we will find in that cell (Figure 4). That effect is compounded by the visitation intensity in the area. The greater the intensity, the greater the number of species threatened for all values of $p_{nature}$ (Figure 4, Table S1). Moreover, the effect of $p_{nature}$ on the number of threatened species is smaller in low intensity regions (first quartile of the $\log_{10}$ nature photo count distribution; Figure 4, Table S1).

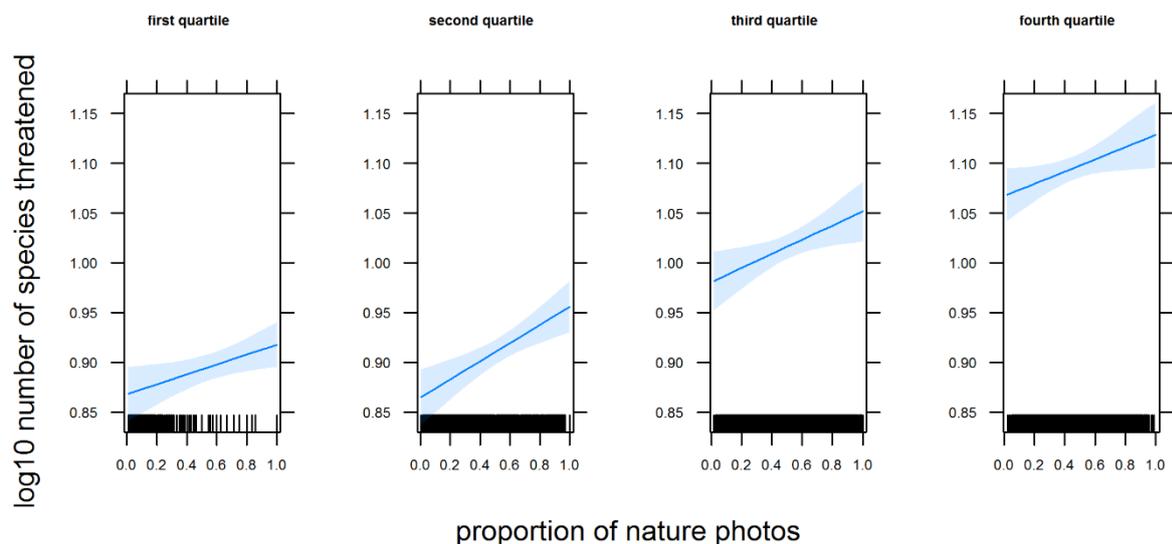

**Figure 4.** Predicted effect of the proportion of nature photos on the number of species threatened by tourism and recreation in each grid cell. Effect was estimated for each quartile of the $\log_{10}$ number of nature photos for each grid cell, a measure of intensity in use of the area by tourism and recreation. Error bands are 95% confidence intervals.

We can therefore use all three variables – number of threatened species, number of nature photos, and proportion of nature photos – in our relative concordance measure (Figure 5) to determine areas that are currently facing a tourism and recreation biodiversity crisis (Figure S2) as well as areas that are currently underused but are likely to receive more pressure as global tourism continues to grow (Figure S3).

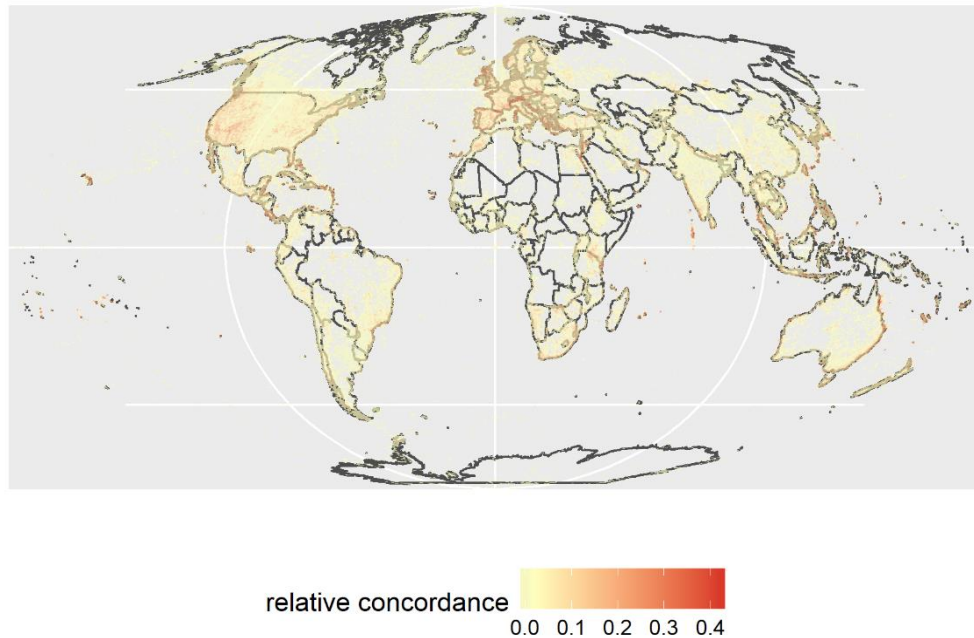

relative concordance

0.0 0.1 0.2 0.3 0.4

**Figure 5.** Relative concordance between the richness in species threatened ($\log_{10}$) by tourism and recreation, the intensity of nature use ($\log_{10}$ number of nature photos), and the focus on nature (proportion of nature photos). All variables were scaled to vary between 0 and 1.

We find that those crisis regions are globally distributed (e.g. Western USA, Caribbean, Maldives, Europe – particularly the Alps, eastern Australia, Philippines, Taiwan) but coastal marine regions emerge as a key threatened feature (Figure S2). Areas that have the scope to face a tourism and recreation threat crisis are essentially neighbouring the current crisis areas (Figure S3), hinting at an edge effect as tourism pressure diffuses from current mature destinations.

**Discussion**

While the biodiversity of both terrestrial (including freshwater) and marine biomes is threatened by tourism and recreation, marine regions affected by these activities will tend to have more species concentrated in threat potential hotspots (Figure 2). The more an area is the focus of human-nature interactions, the more species are threatened by recreation and tourism in this area (Figure 4). This effect is present regardless of how many people use an area but will intensify as the area receives more nature recreationists. We now have regions that are threat potential hotspots and tourism and recreation hotspots too. These regions (Figure 5) require urgent management interventions to address the tourism and recreation biodiversity crisis.  In addition, if we cannot curb tourism pressure in these regions, the spatial extent of the damage is likely to further expand regionally as they tend to be surrounded by areas which have the scope to become crisis areas (threat hotspots but with currently low tourism and recreation pressure).

Our results are of course sensitive to potential sampling biases. The age structure and country of origin profile of tourists and recreationists over the study period tend to match the bias observed in the demographic profile of Flickr users (Balmford et al., 2009; Thomee et al., 2016). Hence, the sampling bias caused by using this social media platform was less likely to bias our results. Recently China has taken over traditional sources of outbound tourism to become the main exporter of tourists. This trend is predicted to further increase in the near future (Gao, Zhang, & Huang, 2018; Lin, Liu, & Song, 2015; Packer, Ballantyne, & Hughes, 2014). This is going to lead to a diversification of tourist attitudes toward nature, which could have positive outcomes for biodiversity conservation (Gao et al., 2018). Future studies must therefore ensure to capture Chinese social media engagements (e.g., Yupoo and WeChat) to understand future global tourism pressures. We also know that all social media platforms have a limited shelf life; mainly driven by the highly competitive nature of the market. If social media sampling is to become a staple approach to understand human behaviour in conservation studies, our community needs to engage early with developers to clearly define and benchmark API features needed that can be ported across platforms as they become available.

The manner to handle this crisis is complex. These analyses show clear trans-national regions where species of wide ranging taxa are affected by intensive recreation from local communities as well as tourism. The situation is particularly worrisome in the Caribbean, Southeast Asia and the Alps. Those regions, and their biomes sensitive to tourism and recreation, are already under pressure from other human impacts which are likely to intensify the effects of tourism and recreation (Heron, Maynard, van Hooidonk, & Eakin, 2016; Pecl et al., 2017). The taxonomic diversity of impacted species means that we do not currently have one conservation intervention that can be effective for all (Bode et al., 2008; van Jaarsveld et al., 1998). We currently do not have one piece of legislation or international regulation that can help us act in a unified manner to solve this crisis (Bramwell & Lane, 2011; Hall, 2011). Indeed, the observed relationship between biodiversity impact and tourism pressure (Figure 4) shows that none of the tourism and recreation governance systems currently employed are successful. We need to develop internationally-fostered trans-national community-led governance of the tourism and recreation biodiversity crisis (Bode et al., 2008; F. Mancini, Coghill, & Lusseau, 2017; Pirotta & Lusseau, 2015). This effort must focus on diverging tourism and recreation pressures, which are set to increase globally, away from threat potential hotspots (Figure 2). We also need to develop new schemes to minimise conservation threats of those activities across taxa, while taking care not to jeopardise the livelihoods of local communities that depend on tourism (F. Mancini et al., 2017; Pirotta & Lusseau, 2015). We have traditionally used spatially-explicit management tools such as PAs to effectively protect biodiversity. PAs are under increasing pressure from a range of human activities (Jones et al., 2018) and generally receive insufficient funding to deal with these impacts. At the same time tourism is increasingly used as a tool to generate income for PA management (Krüger, 2005) and, given the impacts highlighted in this study, this strategy could lead to missing biodiversity targets. Moreover, nature recreation and tourism is increasingly happening outside PAs (Balmford et al., 2009; Pergams & Zaradic, 2008), which highlights the importance of designing new policy tools to manage people-nature interactions outside PAs. We must now re-think the role PAs play as tourism assets if we are to combat the tourism biodiversity crisis.

## Acknowledgements and Data

We would like to thank Jemma Window (Able) from the IUCN Global Species programme for kindly responded to our query and providing the spatial range data for the identified species. We would like to thank IUCN and Yahoo/Flicker for making the respective data available. The YFCC100M dataset is available from Amazon AWS as a S3 data bucket (https://multimediacommons.wordpress.com/yfcc100m-core-dataset/). The IUCN Red List data is available from http://www.iucnredlist.org/. FM is supported by the Dominic Counsell Scottish Natural Heritage studentship.

**Supporting information**

Table S1. Fitted lm-gls models for the effect of the proportion of nature photos in cells on the $\log_{10}$ number of threatened species in those cells for the four quartiles of the total number of photos taken in cells (Figure 4). The residual exponential spatial correlation structure was estimated using latitude and longitude (determined using range and nugget parameters).

| Quartile | intercept | | Effect of $p_{nature}$ | | Residual correlation | |
|---|---|---|---|---|---|---|
| | coefficient (SE) | statistics | coefficient (SE) | statistics | range | nugget |
| 1st | 0.87 (0.055) | $t_{5704}$=15.9, p<0.0001 | 0.050 (0.012) | $t_{5704}$=4.1, p<0.0001 | 9.2 | 0.08 |
| 2nd | 0.86 (0.097) | $t_{6004}$=8.9, p<0.0001 | 0.091 (0.015) | $t_{6004}$=6.0, p<0.0001 | 15.7 | 0.08 |
| 3rd | 0.98 (0.12) | $t_{5911}$=7.8, p<0.0001 | 0.071 (0.019) | $t_{5911}$=3.8, p<0.0001 | 19.8 | 0.10 |
| 4th | 1.07 (0.179) | $t_{5891}$=6.0, p<0.0001 | 0.060 (0.022) | $t_{5891}$=2.8, p=0.005 | 27.7 | 0.09 |

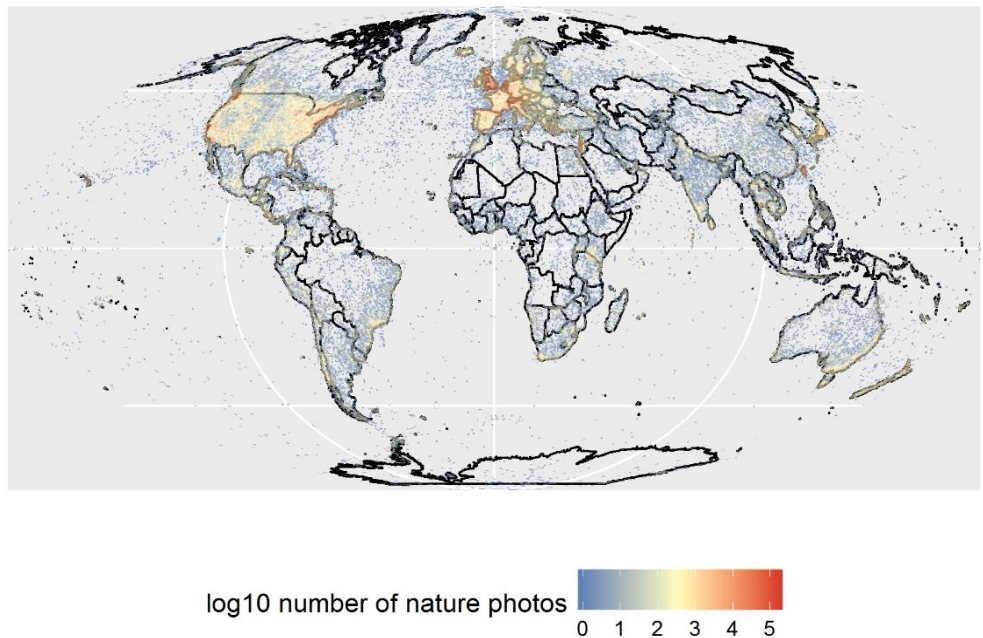

log10 number of nature photos  0 1 2 3 4 5

Figure S1. The number of photos labelled as 'nature' in each 50 x 50 km grid cell

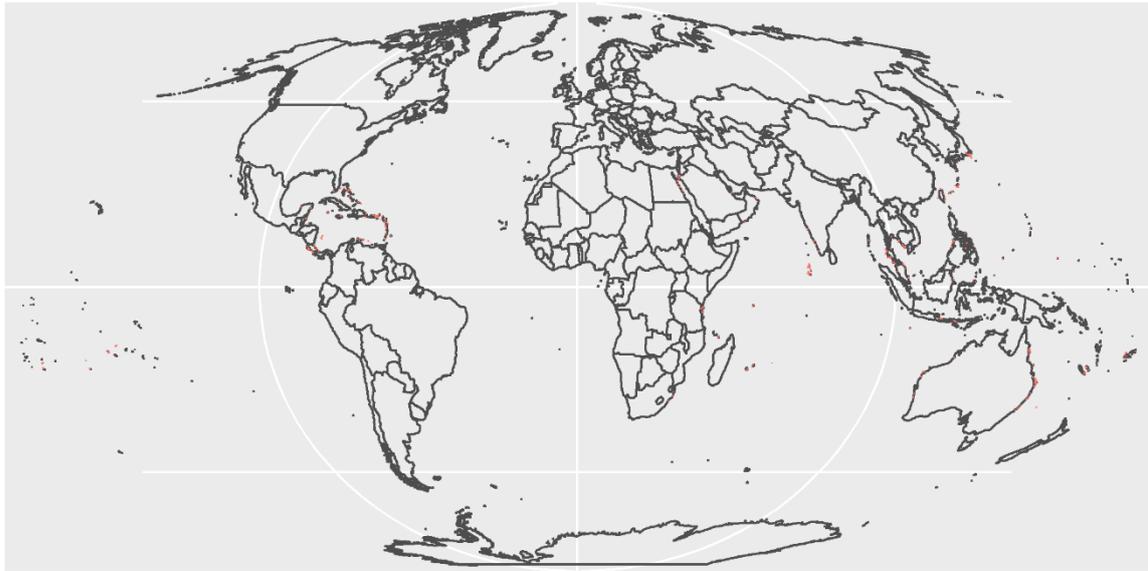

Figure S2. Identified priority areas for crisis management (red): area with high number of threatened species (>100 species), with higher than expected attraction to nature ($p_{nature}$>0.5) and higher use (number of nature photos greater than the 75% quantile)

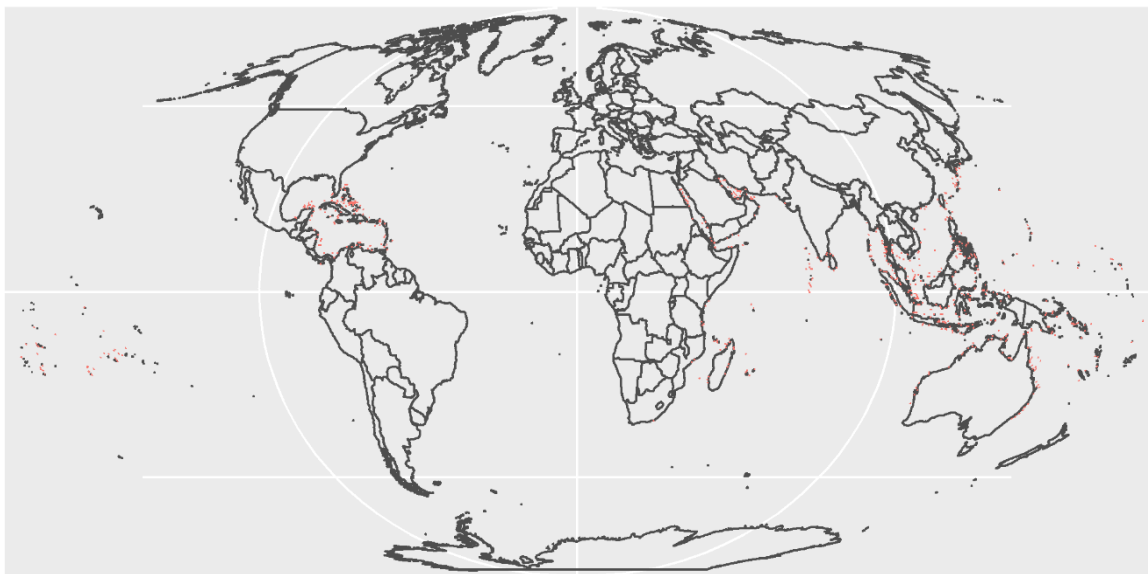

Figure S3. Identified priority areas for preventive management (red): area with high number of threatened species (>100 species), with higher than expected attraction to nature ($p_{nature}$>0.5) and lower use (number of nature photos less than the 25% quantile)